\documentclass[doublecol]{epl2}

\title{Emergence of double  scaling law in complex system}
\shorttitle{Emergence of double  scaling law in complex system} 

\author{D. D. Han\inst{1}\thanks{E-mail: \email{ddhan@ee.ecnu.edu.cn}}\and J. H. Qian\inst{2,3} 
\and Y. G. Ma\inst{2}
}
\shortauthor{D. D. Han \etal}

\institute{
  \inst{1}  School of Information Science and
Technology, East China Normal University, Shanghai 200241, China\\
  \inst{2} Shanghai Institute of Applied Physics, Chinese Academy of Sciences,
Shanghai 201800, China\\
 \inst{3} Graduate School of
  the Chinese Academy of Sciences, Beijing 100080, China}

\pacs{89.75.Hc}{Networks and genealogical trees}
\pacs{89.75.Da}{Systems obeying scaling laws}
\pacs{89.40.Dd}{Air
transportation}

\abstract{We introduce a stochastic model to explain a double
power-law distribution which exhibits two different Paretian
behaviors in the upper and the lower tail and widely exists in
social and economic systems. The model incorporates fitness
consideration and noise fluctuation. We find that if the number of
variables (e.g. the degree of nodes in complex networks or
people's incomes) grows exponentially, normal distributed fitness
coupled with exponentially increasing variable is responsible for
the emergence of the double power-law distribution. Fluctuations
do not change the result qualitatively but contribute to the
second-part scaling exponent. The evolution of Chinese airline
network is taken as an example to show a nice agreement with our
stochastic model. }

\begin{document}

\maketitle

\section{Introduction}

Power law behaviors are now pervasive in various kinds of studies
\cite{hdd,new,BA1,ma,epjb,phone,Ama,Cohen,Cal,Havlin,Lopez,www,qian,Newman},
which give an important class of complex networks, namely the
scale-free networks. However, in some cases, such a single
property is insufficient to describe the distributions in
real-world systems in which scaling law is absent in some regions
or, even more peculiar, changed at some critical points
\cite{cut,cai}. In contrast to the typical power law, distribution
including two different power-law regions is called double
power-law whose cumulative distribution, namely the probability
that variable $K$ is larger than a specific value $k$, is given by
\cite{cai}:

\begin{equation}
P(K>k)  = \{_{k^{-\gamma_2}\\\\,~~k>k_c }^{k^{-\gamma_1}\\\\, ~~
k<k_c; }
\end{equation}
where $\gamma_1$ and $\gamma_2$ are two scaling exponents while
$k_c$ is the turning point. This kind of property exists widely in
social and economic systems such as the degree distribution of
airline network, word network or scientist collaboration network
and the distribution of people's incomes
\cite{cai,Liu,new2,reed,hanhan}.

  Some works related to double power law concentrate on how
to fit such distribution by a uniform function rather than treat
two power separately. Non-extensive statistical theory and the
combination of different power-law functions were applied to the
problem \cite{fitcan,nonextensive}. However these works cannot
tell us how this nontrivial property comes to its being. To
understand its underlying mechanism, Reed proposed a model based
on geometric Brownian motion \cite{reed}. He proved that such
process coupled with exponential distributed evolution time
causes, as he called, a double Pareto-lognormal distribution which
has a lognormal body but power-law behaviour in both tails. This
distribution is shown to provide an excellent fit to observed data
on incomes and earnings. Dorogovtsev and Mendes proposed another
different model aiming to explain the double power-law degree
distribution in word network \cite{dorovge}. The model is
constructed by two mechanisms: the preferential attachment and the
creation of new links between old nodes which increases with
evolution time. By continues approach, they show the degree
distribution has two different scaling exponents, $-1.5$ for upper
tail and $-3$ for lower tail.

  Although the above two models can explain incomes and word network
respectively, both of them have limitations. In word network model
the scaling exponents are fixed. Thus it cannot explain the
distributions with distinct scaling exponents. While in Reed's
model, the relative increase rate of incomes is assumed to be the
same for all people. This is far from our knowledge that persons
have heterogeneous ability of making money. Therefore fitness
character must be taken into account to generalize the model.
Besides, although the previous models reproduced some characters,
the evolution of the real-world systems have not been investigated
to support their model assumptions.

  In this Letter, a general stochastic model is developed to explain
the double power-law distribution. The model incorporates fitness
consideration and noise fluctuation, which is general to describe
many real-world system evolution. We find normal distributed
fitness coupled with exponentially increasing variables is
responsible for the emergence of the double power-law distribution
while fluctuation does not change the result qualitatively but
contribute to the scaling exponent. We also investigate the
evolution of CAN to provide evidence for the proposed model.

\section{Generalized model for double power law}

  Let's denote $k_i(t,t')$  the value at time $t$ of the
$i$-th variable which comes into the system at time $t'$ and
denote $N(t)$ the number of the total variables in the system at
time $t$. Regardless of the specific meaning of $k_i(t,t')$, its
evolution pattern usually shares common features. For an example,
the increasing rate of $k_i$ is proportional to $k_i$ itself. This
is probably caused by the preferential attachment (also called the
rich get richer) that widely exists in self-organized complex
systems \cite{BA1}. Besides, it is natural to assume that the
increasing rate is proportional to some of its own attributes
which is called fitness, denoted as $\eta_i$ \cite{BA2}. In
reality fitness can be interpreted as, for examples, capital,
social skills, activity levels of individuals and population or
Gross Domestic Product (GDP) of cities. The increasing rate can
also be influenced by other ingredients which can be normalized to
be a time-dependent factors. But as the first step, let us focus
on the simplest case where such factors are treated as constants.
In this context, the evolution equation of $k_i(t,t')$ is given
by:
\begin{equation}
\frac{dk_i}{dt} = \eta_ik_i .
\end{equation}
Thus $k_i$ grows exponentially as $k_i(t,t') = e^{\eta_i(t-t')}$
(assuming $k_i(t',t') = 1$). This directly restricts the form of
$N(t)$ in some systems such as node degree evolution in a network.
Limited by its structure, the exponentially growing degree $k_i$
requires the exponentially growing $N(t)$ (or even faster).
Although in some cases such as people's incomes $N(t)$ does not
encounter this problem, it has also been assumed to increase
exponentially \cite{reed}. Therefore we assume
\begin{equation}
N(t) \propto e^{c_nt}.
\end{equation}
The distribution of fitness $\eta_i$ is critical in our model. As
we will analyze, normal distributed $\eta_i$ is essential to
produce the double power-law distribution. Note that when we
choose the specific parameters for $\eta_i$ in a
network-structured system, we meet with the similar problem to
$N(t)$. Since degree of a node can not exceed $n(t)-1$, to keep a
long time evolution $\eta_i$ should be restricted so that the mean
value of $\eta_i$, denoted as $\mu_\eta$, cannot be far larger
than $c_n$ while the standard variance, denoted as $\sigma_\eta$,
should be bounded properly.

  Now let us turn to analyze the distribution under the above-mentioned condition.
Using equation $p(k_i(t,t'))dk_i(t,t') = f(\eta_i)d\eta_i$, the
distribution of $k_i(t,t')$ is easily derived to follow a
lognormal distribution. Since the variables are added
exponentially, their lifetime, defined as $T = t-t'$, follows
exponential distribution. Therefore the value of $k_i$ we actually
examined is lognormal $k_i(T)$ mixed by exponentially distributed
$T$. Thus the distribution reads:
\begin{equation}
p(k) = \int_{0}^{t_c}\frac{1}{\sqrt{2\pi}\sigma_\eta
kT}e^{-\frac{(\ln(k)-\mu_\eta T)^2}{2\sigma_\eta
^2T^2}}e^{-c_nT}dT,
\end{equation}
where $t_c$ is the time when we examine $p(k)$. If
$\sigma_\eta\rightarrow0$, $\frac{1}{\sqrt{2\pi}\sigma_\eta
kT}e^{-\frac{(\ln(k)-\mu_\eta T)^2}{2\sigma_\eta
^2T^2}}\rightarrow\delta(T-\frac{\ln(k)}{\mu_\eta})$ and Eq.(4)
gives:
\begin{equation}
p(k) \propto k^{-(1+\frac{c_n}{\mu_\eta})}\zeta(k-e^{\mu_\eta
t_c}),
\end{equation}
where $\zeta(k) = \{_{0\\\\,~~k>0}^{1\\\\,~~k\leq0}$. The
variables $k_i$ follow a power-law distribution with a cut-off at
$k_c = e^{\mu_\eta t_c}$ which is the maximum value. On the other
hand, $\frac{1}{\sqrt{2\pi}\sigma_\eta
kT}e^{-\frac{(\ln(k)-\mu_\eta T)^2}{2\sigma_\eta
^2T^2}}\rightarrow0$ if $\sigma_\eta\rightarrow\infty$ and the
integral becomes $k$ independent, indicating a uniform
distribution. However this never happens in a network-structured
system since $\sigma_\eta$ is limited as we have discussed.

\begin{figure}
\includegraphics[scale=.65]{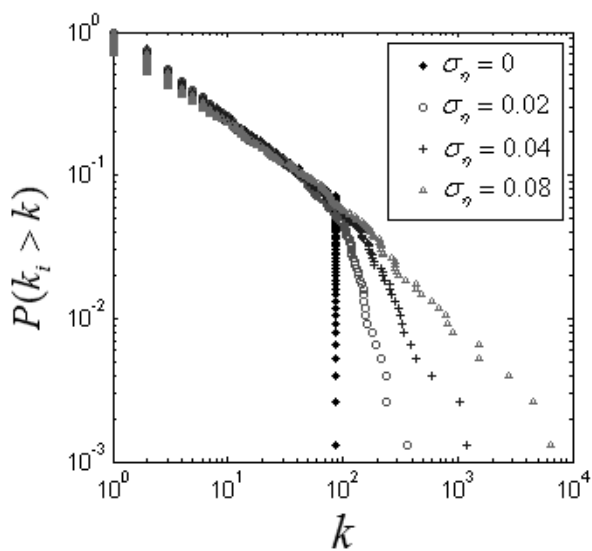}
\caption{\footnotesize Simulation of the cumulative distribution
of variable $k_i$. The simulation is carried out with $\mu_\eta =
0.15$, $N(t) = 50e^{0.088t}$ and $t_c=30$. For other $\mu_\eta$
and exponential increasing $N(t)$, similar results can be
obtained. The study is averaged over $50$ realizations.}
\label{simulation}
\end{figure}

\begin{figure}
\includegraphics[scale=.59]{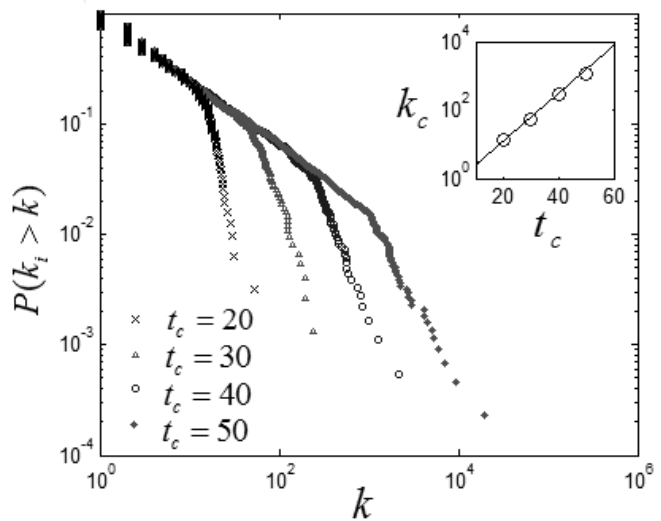}
\caption{\footnotesize Simulation of the cumulative distribution
of $k_i$ at different $t_c$. The simulation is carried out with
$\mu_\eta = 0.15$, $N(t) = 50e^{0.088t}$ and $\sigma_\eta = 0.02$.
Inset: the correlation of $k_c$ and $t_c$. The fitted line (solid
line) represents $y= 0.6e^{0.15x}$, which illustrates that the
turning point $k_c$ increases with $t_c$ as $k_c\sim e^{\mu_\eta
t_c}$.} \label{tc}
\end{figure}

  For a finite $\sigma_\eta>0$, it is difficult to derive analytical result from
Eq.(4). Therefore numerical experiments are applied to analyze the
problem. The simulation is carried out by the following
instructions. At each time step new variables increasing
exponentially are added by initial value 1 and are assigned
fitness chosen from a normal distribution. Then each variable
increases its value according to Eq.(2). We simulate the
cumulative distribution for different $\sigma_\eta$ as shown in
Fig.~\ref{simulation}. When $\sigma_\eta = 0$, the distribution
follows a power-law form with a cut-off at $e^{\mu_\eta t_c}$, as
we have discussed above. With the increase of $\sigma_\eta$, the
second part of the distributions decreases more and more slowly
while their shapes seem to be a power law. It is noteworthy that
the turning point occurs at about $k_c = e^{\mu_\eta t_c}$ which
is exact the point at which the cut-off occurs for $\sigma_\eta =
0$. Therefore the turning point is expected to increase with the
evolution time $t_c$ as $k_c \sim e^{\mu_\eta t_c}$, as well
demonstrated in Fig.~\ref{tc}.

Another interesting discovery is that with the increase of
$\sigma_\eta$, the first part of the distributions does not change
significantly. From Fig.~\ref{simulation}, we see that all the
first part of the curves superpose the distribution of
$\sigma_\eta = 0$ well. Thus the first part of $p(k)$ must follow
the same power-law distribution as the $p(k)$ of $\sigma_\eta = 0$
with exponent $\gamma_1 = \frac{c_n}{\mu_\eta}$ which is
independent of $\sigma_\eta$. However $p(k)$ will finally become
uniform distribution when $\sigma_\eta\rightarrow\infty$. There
must have a transition point $\sigma_{tr}$ at which the first part
of the distribution starts to change. By extensive numerical
experiments, the transition point is determined to be $\sigma_{tr}
\approx \frac{\mu_\eta}{2}$. The distributions of $\sigma_\eta
> \sigma_{tr}$ are not studied since we only concern small
$\sigma_\eta$. When $\sigma_\eta = \sigma_{tr}$, the second part
of the distribution, as seen in Fig.~\ref{simulation}, still
decreases faster than $k^{-\frac{c_n}{\mu_\eta}}$. This result
indicates that for all the $\sigma_\eta \leq \sigma_{tr}$, the
second part of the distributions has an upper bound.

Now let us prove that the second part of $p(k)$ has a lower bound.
It is easy to examine that when $\ln(k)
> (\sigma_\eta\sqrt{\frac{\ln(t)}{t-1}} + \mu_\eta)t$, the following
inequality is valid: \begin{small}
\begin{eqnarray}
\frac{1}{\sqrt{2\pi}\sigma_\eta kT}e^{-\frac{(\ln(k)-\mu_\eta
T)^2}{2\sigma_\eta ^2T^2}}e^{-c_nT}
> \nonumber \\
\frac{1}{\sqrt{2\pi
T}\sigma_\eta k}e^{-\frac{(\ln(k)-\mu_\eta T)^2}{2\sigma_\eta
^2T}}e^{-c_nT}.
\end{eqnarray}
\end{small}
Since $\sigma_\eta$ is usually very small, the above inequality is
approximately considered to be valid when $\ln(k) > \mu_\eta T$.
Therefore for any $k > e^{\mu_\eta t_c}$ (namely the second part
of $p(k)$. The following discussion is restricted to this
condition), the integral of the left term in Eq.(6) from 0 to
$t_c$ must be larger than that of the right term. The integral of
the left term is exactly the degree distribution $p(k)$ while the
integral of the right term, according to Ref.~\cite{www}, follows
asymptotically a power-law function with the exponents related to
$\mu_\eta$, $\sigma_\eta$ and $c_n$. Thus for a specific group of
above exponents, $p(k)$ has a lower bound.

\begin{figure}
\includegraphics[scale=.7]{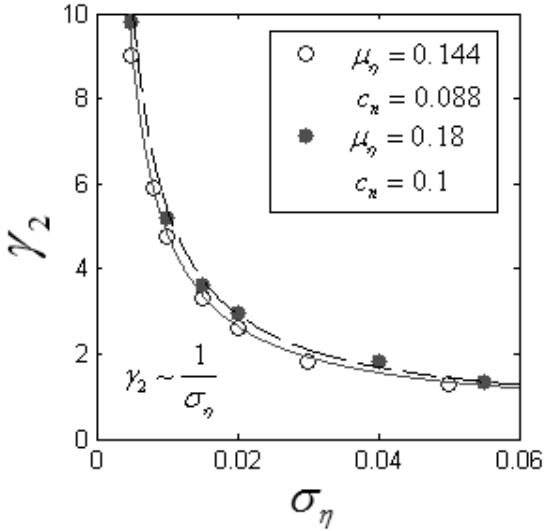}
\caption{\footnotesize Numerical studies of $\gamma_2$. The
evolution time $t_c$ is selected properly to keep the number of
variables at about 1000 while the initial number of variables
equals to 50. The fitted lines (solid line for $\mu_\eta = 0.15$,
$c_n = 0.088$ and dashed line for $\mu_\eta = 0.18$, $c_n = 0.1$)
is given by $\gamma_2 = 0.46 + \frac{0.044}{\sigma_\eta}$ and
$\gamma_2 = 0.44 + \frac{0.05}{\sigma_\eta}$, respectively. The
study is averaged over 50 realizations. For other $\mu_\eta$ and
$c_n$, similar results can be obtained.} \label{sigama}
\end{figure}

  If $p(k)$ is not oscillatory, the existence of the upper and the lower
bound allows $\frac{\ln(p(k))}{\ln(k)}$ to have a limitation.
Assuming it to be $-\gamma$, then $p(k)$ is written as $p(k)
\propto l(k)k^{-\gamma}$, where $l(k) \sim o(k^\beta)$ and
$k^{-\beta} \sim o(l(k))$ are valid for any $\beta > 0$ when $k
\rightarrow \infty$. Therefore for large $k$, given any constant
$u > 1$, we have the following inequality:
\begin{equation}
\frac{(uk)^{-\beta}}{k^{-\beta}} <
\frac{l(uk)}{l(k)}<\frac{(uk)^\beta}{k^\beta}.
\end{equation}
Let $\beta \rightarrow 0$, then we have
\begin{equation}
\lim_{k \rightarrow \infty}\frac{l(uk)}{l(k)} = 1.
\end{equation}
Note that Eq.(8) is also valid for any constant $0< u \leq1$. This
property of $l(k)$ follows directly from the requirement that
$p(k)$ is asymptotically scale invariant. Thus, the form of $l(k)$
only controls the finite extent of the lower tail and will not
affect its scaling exponent significantly. So the second part of
the $p(k)$ is also power law. For further study, numerical
simulation is applied to determine the exponent of the second part
of $p(k)$, denoted as $\gamma_2$. It is found that
\begin{equation}
\gamma_2 \sim \frac{1}{\sigma_\eta}.
\end{equation}
The result is well demonstrated by the simulation as shown in
Fig.~\ref{sigama}. It is noteworthy that $\sigma_\eta \rightarrow
0$ leads to $\gamma_2 \rightarrow \infty$. The second part of
$p(k)$ degenerates naturally to be a cut-off. The exact
formulation of $\gamma_2$ may also be related to $c_n$ and
$\mu_\eta$, but extensive simulations indicate that $\gamma_2$ is
much less sensitive to $c_n$(or $\mu_\eta$) than to parameter
$\sigma_\eta$.

\begin{figure}
\includegraphics[scale=.55]{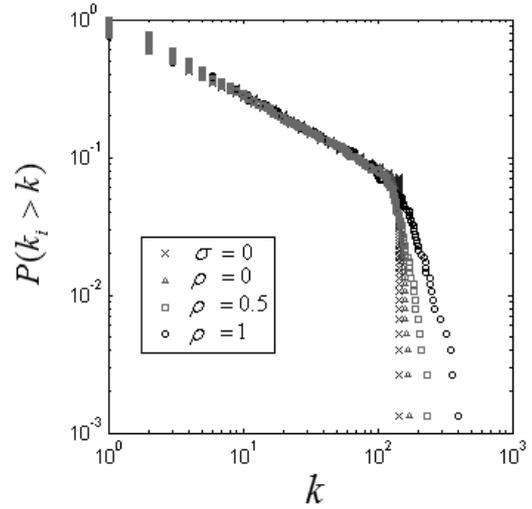}
\caption{\footnotesize Numerical studies of the cumulative
distribution for different $\rho$. The simulation is carried out
with $\mu_\eta = 0.17$, $N(t) = 50e^{0.088t}$ and $t_c=30$.
'$\times$' represents $\sigma_\eta = \sigma=0$ and the others
represent $\sigma_\eta = \sigma  = 0.02$. When $\sigma_\eta =
\sigma = 0$ the distribution is a power law independent of $\rho$.
When $\sigma_\eta = \sigma > 0$, the first-part scaling exponent
does not change while the second-part exponent decreases with
$\rho$. The study is averaged over 50 realizations.} \label{rhoo}
\end{figure}

  So far we have provided a possible model to produce double
  power-law distribution. However real complex systems such as the Internet
or WWW usually include fluctuations that may be essential to
describe the dynamics of its evolution \cite{www,www2}. Therefore,
a general model must be able to describe this feature. The
generalization can be carried out by modifying Eq.(2):
\begin{equation}
\ d{k_i} = \rho\eta_ik_idt + (1-\rho)(\mu_\eta dt + \sigma dw)k_i,
\end{equation}
where $dw$ is white noise of standard normal distribution and
$\sigma$ is the standard variance of fluctuations. $\rho\in[0,1]$
is a parameter representing the relative contribution of the noise
and the fitness, which can be considered as a measure of the
degree of the disorder in real-world systems. Note that if
$\rho=1$, Eq.(10) becomes Eq.(2) while if $\rho=0$, it leads to
the geometric Brownian motion. Fluctuation described by the term
$\sigma k_idw$ is based on a general fact that in real systems
such as WWW, site with large number of connections are likely to
lose or gain more links than the site with small one. The solution
of Eq.(10) is solved to be
$e^{\rho\eta_it}e^{(1-\rho)[(\mu_\eta-0.5\sigma^2)t+\sigma w]}$
which follows lognormal distribution with logarithmic mean
$\rho\mu_\eta t+(1-\rho)(\mu_\eta-0.5\sigma^2)t$ and logarithmic
variance $(\rho\sigma_\eta t)^2+(1-\rho)^2\sigma^2t$. By the
similar methods used above, one can verify the distribution is
still a double power law but the second scaling exponent is
controlled by parameter $\rho$. In Fig.~\ref{rhoo} we show the
cumulative distribution for different $\rho$. It is found that the
first part power-law behavior is still independent of $\rho$,
leading to $\gamma_1 = \frac{c_n}{\mu_\eta}$, but the second
power-law exponent decreases with $\rho$. Therefore noise
fluctuations do not change the distribution qualitatively but
contribute to the second scaling exponent.

  The present model (Eq.(10)) indicates that evolution of a complex system
may be characterized by two parts: a leading ingredient
influencing the evolution and noise fluctuations. This will be
interpreted as follow. Since there are usually various ingredients
related to the evolution of complex system, practically we cannot
take all of them into account. A feasible method is to consider
the most important ingredient as the fitness while all other minor
ones as contributions to fluctuations. Then parameter $\rho$
represents how much the leading ingredient contributes to the
evolution. If the evolution is totally governed by the leading
ingredient, then $\rho \rightarrow 1$, indicating a deterministic
pattern. On the other hand, if there is no apparent leading
ingredient, it leads to $\rho \rightarrow 0$, indicating a random
picture. Therefore our model is general to describe various
real-world systems which evolve between order and disorder, and
provide a better understanding on their evolution. As we will see
in the following section, CAN is a typical example that follows
such an evolution mechanism.

\section{An example: Evolution of CAN}

  In this section, we will analyze the evolution of CAN to provide
evidence for our proposed model.

  Chinese airline system can be modeled as a complex network with cities
representing nodes and flights representing edges. The degree of a
node is defined as the sum of the airlines connecting to it. The
degree distribution of CAN has been investigated by several
studies which all indicate a double power-law behavior
\cite{cai,Liu,fitcan}. Here we will report some useful
information. First we have analyzed the total number of nodes
$N(t)$ existing at time $t$. It grows as $N(t) \propto e^{c_nt}$
with $c_n \approx 0.088$, which is consistent with our assumption
that nodes increases exponentially. The number of edges $M(t)$
also increases exponentially with time as $M(t) \propto e^{c_mt}$
with $c_m \approx 0.154$. We have also measured the parameters of
the degree distribution of CAN from 1999 to 2003 and a single year
2008, summarized in Table.~\ref{pk}. The exponent $\gamma_1$
stabilizes at about 0.51 while $\gamma_2$ fluctuates from 2.1 to
2.7. Note that the stabilization of $\gamma_1$ is indicated by our
model where $\gamma_1$ is independent of fluctuations. The turning
point $k_c$ shows an increase from 18 to 30, which is also
consistent with our result that the turning point increases with
evolution time.

\begin{figure}
\includegraphics[scale=.5]{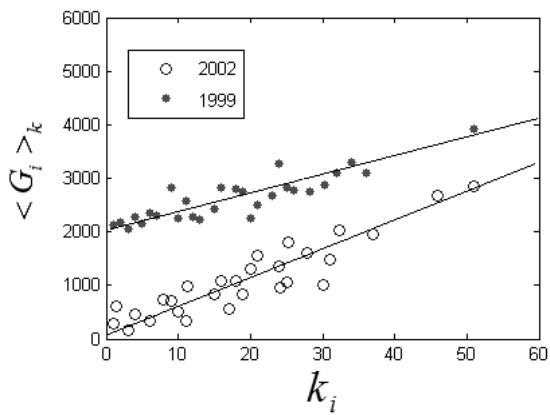}
\caption{\footnotesize The correlation of degree and its
corresponding GDP for year 1999 and year 2002. Both of them
exhibit linear correlation. $\langle G_i(t)\rangle_k$ is an
average over all the nodes with the same degree $k$. Note that the
data of the two years are not drawn for better visualization.
($G_i(1999)$ here increases 2000 from the origin data.) The
correlation does not change from 1998 to 2003.}
\label{correlation}
\end{figure}

In the present paper the evolution of CAN is studied by
investigating the correlation between GDP  and degree. The
economic growth such as the size of tertiary industry has been
recently demonstrated to be a leading ingredient in shaping the
topology of CAN \cite{liu2}. According to the discussion in the
last section, we consider GDP relates to the fitness of the
corresponding nodes. Note that the evolution of degree can also be
studied by directly measuring the logarithmic ratio of the degree
of successive two years. But this method can neither help to
distinguish the identity of the fitness nor provide useful
information of the corresponding parameters which is important in
our analysis.  As shown in Fig.~\ref{correlation}, we found that
the degree forms a linear function with its corresponding GDP
($R^2 > 0.62$) while the fluctuations are obvious. Despite the
continuing evolution of both GDP and degree, this correlation has
maintained for at least six years ($1998-2003$) since it has been
first observed in $1998$. Therefore it provides some key
information about the evolution of degree in CAN and cannot be
viewed as just a coincidence.

Considering the time evolution, the correlation can be described
as:
\begin{equation}
k_i(t) = D(t)G_i(t),
\end{equation}
where $G_i(t)$ is the GDP of city $i$ at year $t$. It grows
exponentially as $G_i(t)\propto e^{\lambda_i t}$ where $\lambda_i$
follows normal distribution with the mean of $0.18$ and the
standard variance of 0.02. The strong positive correlation
confirms that economy may govern the evolution of the degree in
CAN while the fluctuations, as we mentioned previously, are
considered to result from some minor ingredients (such as
population density, public administration, geographical
constraints, etc). Both the two aspects contribute to $D(t)$,
leading to an expression given by $D(t) =
e^{a(t)}e^{\varepsilon(t)}$, where term $e^{a(t)}$ is the
time-dependent slope and the term $e^{\varepsilon(t)}$ represents
the fluctuations.
\begin{table}[t]
 \centering
 \caption{Two scaling exponents ($\gamma_1$ and $\gamma_2$) and the turning point ($k_c$) of cumulative degree distribution in CAN}\label{pk}
 \begin{tabular}{c c c c c c c}
     \hline
       \multicolumn{1}{c}{}&\multicolumn{1}{c}{1999}&\multicolumn{1}{c}{2000}&\multicolumn{1}{c}{2001}&\multicolumn{1}{c}{2002}&\multicolumn{1}{c}{2003}&\multicolumn{1}{c}{2008}\\
   \hline
   $\gamma_1$&0.46 &0.51& 0.52& 0.51& 0.51& 0.51 \\
   $\gamma_2$&2.2 &2.1& 2.5& 2.3& 2.7& 2.7 \\
   $k_c$&18 & 18& 20& 21& 22& 30 \\
     \hline
 \end{tabular}
\end{table}

\begin{figure}
\includegraphics[scale=.48]{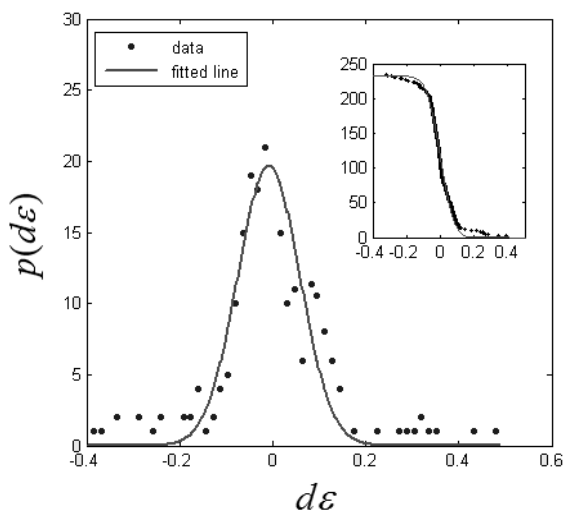}
\caption{\footnotesize The distribution of the increment of
$\varepsilon(t)$ for all the five years. The data is binned into
classes. The fitted line is $p(d\varepsilon) =
19.4e^{(-\frac{(d\varepsilon)^2}{2*0.09^2})}$. Therefore it
follows normal distribution with mean 0 and standard variance
0.09. Inset: The cumulative distribution of $d\varepsilon(t)$. It
is well fitted by
$P(d\varepsilon(t)>d\varepsilon)=116erfc(\frac{d\varepsilon}{0.127})$
where $erfc(x)=\frac{2}{\sqrt{\pi}}\int_x ^\infty e^{-x^2}dx$.}
\label{zengliang}
\end{figure}
The specific form of $a(t)$ is easy to be evaluated. Summing
Eq.(11) for all nodes we get $\langle D(t) \rangle = e^{a(t)} =
\frac{2M(t)}{\sum_i G_i(t)}\sim e^{-0.036t}$, where $\sum_i
G_i(t)$ is measured to be proportional to $e^{0.19t}$. Thus $a(t)
= -0.036t$. Then $\varepsilon(t)$ can be investigated from data by
calculating $\varepsilon(t)=\ln(\frac{k_i(t)}{G_i(t)})-a(t)$.
However what we concern here is the increment of $\varepsilon(t)$,
defined as $d\varepsilon(t) = \varepsilon(t)-\varepsilon(t-1)$. In
Fig.~\ref{zengliang} we plot the distribution of $d\varepsilon(t)$
for all the five years. It follows a normal distribution with the
mean of 0 and the standard variance of 0.09. Furthermore we
calculate the self-correlation function of $d\varepsilon(t)$,
defined as $\langle d\varepsilon(t)d\varepsilon(t+\tau)\rangle =
\frac{1}{N(t)}\sum_i d\varepsilon_i(t)d\varepsilon_i(t+\tau)$
($\tau$ is the time interval). As listed in Table.~\ref{de}, it
exhibits very weak correlation (correlation coefficient $<0.1$)
when $\tau\neq0$. Therefore $d\varepsilon(t)$ can be regarded as
white noise and expressed as $d\varepsilon(t)=0.09dw$. Then $D(t)$
is written as $D(t) \sim e^{0.09w-0.036t}$. Substituting it into
Eq.(11) and applying differentiation we have \cite{jieshi}
\begin{equation}
dk_i = (\lambda_i-0.036)k_i dt + 0.09k_i dw.
\end{equation}
Eq.(12) is exact the form of our model which can give rise to
double power-law distribution. To further demonstrate its
agreement with the real evolution of CAN, we simulate the degree
distribution according to Eq.(12). We obtained $\gamma_1 = 0.61$
(it can also be calculated from $\gamma_1 = \frac{c_n}{\mu_\eta} =
\frac{c_n}{\langle\lambda_i\rangle-0.036} = 0.61$), comparable to
the first exponent 0.51 while $\gamma_2 = 2.84$, good agreement
with the second exponent 2.7.

\begin{table}[t]
 \centering
 \caption{The self-correlation function $\langle d\varepsilon(t)d\varepsilon(t+\tau)\rangle$}\label{de}
 \begin{tabular}{c c c c c c}
     \hline
       \multicolumn{1}{c}{}&\multicolumn{1}{c}{1999}&\multicolumn{1}{c}{2000}&\multicolumn{1}{c}{2001}&\multicolumn{1}{c}{2002}&\multicolumn{1}{c}{2003}\\
   \hline
   $1999$&1&0.041&-0.019& -0.02& -0.01\\
   $2000$&0.041&1 &-0.06& -0.023& 0.06\\
   $2001$&-0.019&-0.06 &1& -0.1& 0.01\\
   $2002$&-0.02&-0.023 & -0.1& 1& -0.035\\
   $2003$&-0.01&0.06 &0.01& -0.035& 1\\
     \hline
 \end{tabular}
\end{table}

\section{Conclusion}
   We have proposed a general model to explain the
emergence of the double power-law distribution. The model
incorporates  fitness consideration and noise fluctuation which
indicates that evolution of a complex system may be characterized
by two parts: a leading ingredient and noise fluctuations. We find
that normal distributed fitness coupled with exponentially
increasing variables is responsible for the emergence of the
double power-law distribution. Fluctuations do not change the
result qualitatively but contribute to the value of scaling
exponent. We have also studied empirically the CAN which turns out
to follow the same evolution pattern as our proposed model.

We have only discussed the behavior of our model when $\sigma_\eta
<\frac{\mu_\eta}{2}$. If $\sigma_\eta$ is not much larger than
$\frac{\mu_\eta}{2}$, the distribution still decays like a double
power-law but both the exponents are different from previous ones.
With the continuing increasing of $\sigma_\eta$, the double
power-law behavior turns out to be unconspicuous. It results from
that the second scaling exponent is gradually close to the first
one and finally becomes indistinguishable.

Finally, we would like to mention that we have done tests for six
usual distributions, namely exponential distribution, uniform
distribution, power-law distribution, Poisson distribution,
Rayleigh distribution and Weibull distribution for the fitness
instead of normal distribution, the results show that none of them
is able to achieve double power-law distribution. Therefore we
believe  that normal distribution of fitness is a key ingredient
responsible for the double power-law distribution.

\vspace{2cm}


\section{Acknowledgements}

This work was partially supported by National Nature Science
Foundation of China under Grant numbers 11075057, 11035009 and
10979074, and the Shanghai Development Foundation for Science and
Technology under contract No. 09JC1416800.


{}


\begin{thebibliography}{}
\bibitem{jieshi} Actually $dw$ as well as $k_i$ defined here is time-discrete,
therefore Eq.(12) is better to be expressed as a difference
equation. However we omit rigor, for simplicity.
\end{thebibliography}
\end{document}